# NEW PERSPECTIVES TO REDUCE STRESS THROUGH DIGITAL HUMOR

**Misnal Munir[1], Amaliyah[2], Moses Glorino Rumambo Pandin[3]**

Universitas Gadjah Mada, Yogyakarta, Indonesia[1,] Universitas Trilogi, Jakarta, Indonesia[2]
Universitas Airlangga, Surabaya, Indonesia[3]

misnalmunir@ugm.ac.id[1], amaliyah@trilogi.ac.id[2], moses.glorino@fib.unair.ac.id[3]

## ABSTRACT

This study aimed to find new perspectives on the use of humor through digital media. A qualitative approach was used to conduct this study, where data were collected through literature review. Stress is caused by the inability of a person to adapt between desires and reality. All forms of stress are basically caused by a lack of understanding of human's own limitations. Inability to fight limitations that will cause frustration, conflict, anxiety and guilt. Too much stress can threaten a person's ability to deal with the environment. As a result, employees develop various kinds of stress symptoms that can interfere with their work performance. Thus, the management of work stress is important to do, one of which uses humor. However, in the digital age the spread of humor can be easily facilitated. The results of this review article find new perspectives to reduce stress through digital humor, namely interactive humor, funny photos, manipulations, phanimation, celebrity soundboards, and powerpoint humor. The research shows that the use of humor as a coping strategy is able to predict positive affect and well-being work related. Moreover, digital humor which has various forms as well as easy, fast and wide spread, then the effect is felt increasingly significant

## 1. Introduction

According to Hariyono (2004) work stress can be experienced primarily by someone, if the person does not have qualified abilities, both technical and mental abilities to cope with the demands of work given to him. The perception that arises from each task that is received and what must be done is that the task is very heavy; lack of resources needed to carry out the duties and responsibilities imposed; or do not have enough ability to be able to achieve the expected results. When such feelings arise in a person, it can be said that the person is experiencing work stress.

Various efforts have been taken to be able to manage, reduce or avoid stress, one of which is by enjoying humor (Reyes, 2012). Perception that arises every time you hear the term humor can not be separated from something that is considered funny, fun, and entertaining. Besides having these characteristics, humor also actually has a positive impact that is able to encourage the emergence of positive emotions (Collum et al., 2011) and improve human health (Samson et al., 2008).

Humor is an easy alternative to be obtained by people who want to get away from workloads or life burdens, so Torreta (2014) states that humor is mechanism coping a powerful to reduce work stress. Regarding the relationship between humor and work stress, Collum et al. (2011) in his research found that the use of humorous comedy videos can actually reduce the level of anxiety and stress in workers. Wijaya (2017) found specifically that superiority and self-defeating humor actually increase stress, whilehumor affiliative can reduce stress.

Along with the development of communication and information technology, then humor that was previously spread through conventional means, such as with television intermediaries, in stage shows, or in forums that are formal or informal, so now humor can also be disseminated through the media on line. The digital media makes it easier for each individual to spread and accept humor in various forms, both in the form of interactive humor, funny photos, manipulations, phanimation, celebrity soundboards, and powerpoint humor (Shifman, 2007). The interesting thing is that there are not many previous researchers who have conducted a direct study of the influence of humor spread with digital media, or hereinafter referred to as digital humor, on work stress. Most researchers still study humor in its original form without considering its distribution media and still focus on certain forms of humor, such as Techarungroj & Mueang Jamnong (2014) who research about internet memes and their ability to spread in the digital world.

## 2. Method
This study is a qualitative literature review in which a researcher was allowed to analyze and evaluate both quantitative and qualitative literature within a domain to draw conclusions about the state of the field.

## 3. Result and Discussion
### 3.1 Work Stress
Robbins & Judge (2015) states that work stress arises as a result of conflict between individuals, both physically and psychologically, with the situation or problems encountered in carrying out their work activities. The situation or problem can be related to work demands, opportunities that must be utilized, or related to the condition of resources needed to carry out productive activities. Gibson et al. (1993) states the same thing, that work stress is a form of reaction to the individual that arises because of demands from the work environment that do not have compatibility with the physical or mental capabilities of the individual.

Mangkunegara (2008) mentions specifically, that work stress is a lot experienced by employees who work at a company, which has an impact, both directly and indirectly, on the attitudes, feelings, thoughts, and behavior of these employees. The higher the work stress will cause the more unstable feelings or emotions, the more negative attitudes and behaviors, and the tense thoughts that are full of anxiety. According to Hariyono (2004), work stress arises because of inability to overcome the situation or problems encountered at work. Employees who experience work stress will become less productive and tend to not be able to develop properly in accordance with the demands of the company. The employee will even have difficulty interacting with various important aspects of his work, including with coworkers, with his supervisor or supervisor, or with relationships or customers of the company.

Job stress can appear in a person due to certain factors. According to Robbins (2012), there are at least three main factors that cause the emergence of work stress, namely:
1. Environmental Factors can include factors that are unstable regional economy, chaotic political conditions, and technological uncertainty.
2. Organizational factors include task demands, position/role demands, individual demands, organizational structure, organizational leadership, and the stage of organizational development.
3. Individual factors include a variety of family problems, economic problems, and are related to individual characteristics or personalities.

Work stress can be measured using two dimensions, namely (Cohen & Williamson, 1988):
1. Perceived helplessness is a feeling that an individual has that he has no control over the surrounding environment, so that the individual is in a state of being uncomfortable, not motivated, and emotional.

2. Perceived self efficacy is a feeling that is the opposite of perceived helplessness in which individuals have strong beliefs about the ability of oneself to do things desired to achieve certain goals.

**3.2 Humor**
Humor is a term often used in everyday life to illustrate some things that are funny and entertaining. The definition of humor mentioned by Reyes et al. (2012) as "the presence of amusing effects, such as laughter or well-being sensations", that is, humor is something that can create confusing effects, which can make people laugh or feel happy sensations. A similar definition of humor was stated by Martin (2003), that humor is a construct that has many dimensions, which mainly has the ability to present feelings of pleasure to oneself and others. Humor can come from a certain pattern of behavior or attitude from someone, which is generally used to build social relationships.

The definition of humor from a different point of view was put forward by Siswanto (2007), that humor is the nature of individuals in seeing things from a non-serious side, where the problems they face are sometimes considered to be funny, so as not to create self-burdening pressures. If Siswanto (2007) and Martin (2003) define humor based on the affective side, then Colom et al. (2011) defines humor from the cognitive side, that humor is actually a thing that has a close relationship with the cognitive aspects of individuals, namely the ability to think or ways of thinking that look at things from an unusual perspective. If individuals generally view a problem as something that must be seriously considered, then the humor that is owned by the individual will direct the perspective that considers the problem as unique, funny, even interesting to face and overcome.

Collum et al. (2011) also states that humor includes all the stimuli that can cause laughter, both in the form of stimulus games, funny stories, cartoons, and others. Laughter referred to in this case can arise due to a sense of superiority, incompatibility, accompanied by the release of emotional or cognitive stress. Samson et al. (2008) added, that humor has the main function to vent or release emotions, sentiments, or other feelings that in general can have a positive influence on human health. Humor is not always the main thing that someone needs to be able to get a relaxed feeling, however, it has become a common thing to know that humor can make you feel tense or depressed, either because of problems experienced in everyday life or because of workloads, become lighter. Related to this, Torretta (2014) states that "humor is often regarded as one of the highest forms of coping with life stress", meaning, humor is often considered as a form of way to cope with life stresses.

According to Raskin (1985), humor can be formed from three sources, namely incongruity, arousal-safety, and disparagement.
1. Incongruity is the juxtaposition of the two or more incongruous parts roomates or circumstances led to humor. Incongruity or interpreted as oddities, inaccuracies, or mismatches between two or more things that are not appropriate but are aligned. The discrepancy creates a strange perception that tends to be funny which is called humor.
2. Arousal-Safety is the humor based on an escape of some form. Safety is an important thing that every individual wants to feel. When an individual's safety is disturbed or threatened, but it turns out later the individual managed to survive, then sometimes the situation can be a source of humor based on how to avoid being threatened.
3. Disparagement is the humor that is the result of hostility, superiority, malice, aggression, derision, or disparagement. Disparagement is humor that is the result of hostility, superiority, malice, aggression, ridicule, or humiliation. In this case, humor is created when different perspectives and interpretations are used to think about and respond to various situations or situations that tend to be negative, so perceptions that are generally negative can be changed to be positive through the creation of humor.

According to Catanescu & Tom (2001), humor can be divided into seven, namely:
1. Comparison is a type of humor created by putting two or more elements together to form a funny situation.
2. Personification is a type of humor created by linking human characteristics with animals, plants, or other objects.
3. Exaggeration is a type of humor created by exaggerating things than they should.
4. Pun is a type of humor created by using language elements to create new meanings.
5. Sarcasm is a type of humor that is created from an ironic response from a situation openly.
6. Silliness is a type of humor created by making funny expressions for ridiculous situations.
7. Surprise is a type of humor that is created from an unexpected situation.

### 3.3 Digital Humor
In accordance with the rapid development of communication and information technology, the humor that was originally created and delivered through conventional means, such as being staged on the stage, aired on television stations, or delivered in direct interaction between individuals, both in formal and formal spheres non formal, at this time can also be conveyed using various types of online media, so that people are increasingly easy to be able to accept various types of humor and consume it as a means to release tension or stress due to various problems encountered in daily life or at work. The digital world in addition to making the spread of humor easier and wider, also causes the development of forms of humor to become more varied. According to Shifman (2007), digital humor can have six forms of presentation, namely interactive humor, funny photos, manipulations, phanimation, celebrity soundboards, and powerpoint humor.
1. Interactive Humor is the form of text contains funny words that require active participation from the recipient of humor to do certain things than just reading, listening, or seeing.
2. Funny Photos is a photo that displays funny messages, which are often presented with funny writing that provides additional description.
3. Maniphotos are photos that are manipulated by combining them with other photos so as to create a funny or strange impression.
4. Phanimation is a moving or animated version of maniphotos.
5. Celebrity Soundboards are digital collections of footage from film and or sound scenes that appear on television or radio owned by actors, which are deliberately cited in the form of sound pieces or short videos that contain funny or strange messages.
6. Powerpoint Humor is funny text or images that are presented in the form of presentations PowerPoint.

### 3.4 Humor Model
Kuiper (2012) has the concept that a sense of humor as a characteristic of differences in individual diversity involving four main styles, namely, affiliative, self-enhancing, aggressive, and self-defeating humor. Both the humor style affiliative and self-enhancing generally touch the positive or adaptive aspects of the sense of humor; while the aggressive and self-defeating style generally touches on the negative or maladaptive aspects of the person.

Martin et al. (2003) & Oktug (2017) further explain the four characteristics of humor, including:
1. Affiliative is humor or a joke created to encourage increased relations between individuals. The characteristics of humor include non-attacking, tolerant, cheerful, containing positive emotions, and maintaining self-esteem.
2. Self-enhancing is humor that aims to defend themselves to avoid negative or non-conducive situations that have the potential to harm oneself. The characteristics of humor include being open, maintaining self-esteem, psychologically healthy, and focusing on internal psychic aspects.
3. Aggressive is humor that is done without regard to its impact on others by saying funny words that actually have the potential to hurt or hurt the feelings of others. These humorous characteristics include sarcasm, teasing, mocking, demeaning, and insulting. This humor also has a close relationship with situations full of anger, aggression, danger, and neuroticism.
4. Self-defeating is humor that is done by humbling oneself to create jokes for others. This humor has the characteristics of a form of defensive or refusal to cover up the negative feelings they have. Humor deals with emotional needs, avoidance, low self-esteem, and anxiety.

**3.5 Coping Stress Through Humor**
Individual efforts in dealing with stress are commonly known as coping. Understanding coping is an effort to manage stressful situations regardless of the results of these efforts (Lazarus & Folkman, in Martin, 2007). This means that the strategies carried out by individuals cannot be considered better than other individuals. The effectiveness of a coping strategy is only determined by its impact in specific situations and its long-term impact. There are many ways to coping with existing stress, both those that focus on the problem, emotions, or how to assess a condition. The means used for coping also vary, one of which is coping stress using humor.

Some proverbs that are widely known states that "laughter is the best medicine" or "laughter is the best medicine". According to Markman (2017), a cognitive scientist from the University of Texas, humor can influence the way a person sees problems and reduces stress experienced. The view that humor has positive benefits in dealing with stress is in line with the results of several studies that have been conducted, which show that individuals with good sense of humor and using it as a coping strategy will be better able to deal with stresses that hit and adjust (Wu & Chan, 2013; Overholser, in Martin, 2007).

In the work context, research related to the use of humor as a coping strategy has been carried out by Doosje et al (2010). This research was conducted on 2094 male and female employees in the Netherlands using the measurement tool Questionnaire of Occupational Humorous Coping (QOHC). The research shows that the use of humor as a coping strategy is able to predict positive affect and well-being work related.

In addition, research on teaching staff at Cukurova University, Turkey also shows that in work, humor used as a coping strategy is associated with lower burnout rates (Tumkaya, 2007). Furthermore Colom et al (2011) conducted a research to develop strategies to reduce levels of anxiety and stress by using positive humor media. This type of quantitative research with a sample of 31 people showed that anxiety and stress levels decreased significantly after seeing comedy videos that contained positive humor.

A study conducted by Widyowati & Priambodo (2016) with the title "Relationship between humor sensitivity and work stress in agricultural quarantine class 1 employees in Semarang, Central Java" aims to determine the relationship between humor sensitivity and work stress in employees of Agriculture Quarantine Class 1 Semarang Java The middle. With quantitative methods and has a sample size of 52 people, showing the results that the higher the sensitivity of humor that employees have, the lower the work stress of employees. The examples above prove that in the context of work, the use of humor can produce a variety of positive effects on the psychological condition of an individual related to work, and corroborate the expert statements listed earlier in this study.

However, the use of humor as a coping strategy is not always beneficial. According to Markman (2017), improper use of humor, such as making oneself or another person a joke can make others have a negative view and reduce social support, which can impact on higher stress levels. In addition, several studies have shown that the use of humor as a coping strategy is an ineffective way to deal with stress, so that its use when compared with other coping strategies includes low frequency (Gutierrez, et al, 2007; Washburn-Ormachea, et al, 2004; Wu & Chan, 2013).

A study conducted by Rahayu & Hadrianmi (2015) examined the relationship between sense of humor and stress on "SLB C" teachers. Through a quantitative approach with a sample of 30 people, obtained results that the sense of humor has no relationship with stress. Whereas Wijaya (2017) examines the role of humor by mediating from subjective well being to stress reduction in early adulthood. This study was quantitative in nature with a total sample of 200 people. The results showed that subjective well being mediated the effect of humor on stress. Humor has a significant negative effect on subjective well being. Humor kind of superiority and self-defeating actually increase stress, while the kind of humor affiliative can reduce stress.

In addition, the use of humor as a coping strategy is also not universal. For example, in China, humor is seen as something less honorable. This is due to the culture adopted, so individuals must continue to maintain behavior in accordance with polite and polite ethics (Yue, 2010). The data above is part of the results of research that is not in line with the results of the study which shows the positive impact of using humor as a coping strategy.

## 4. Conclusion

A review of previous studies produced findings that humor has a positive influence on work stress levels. The higher the humor sensitivity that employees have, the lower the work stress of Widyowati & Priambodo employees (2016). Doosje et al (2010) research shows that the use of humor as a coping strategy is able to predict positive affect and well-being work related. Moreover, digital humor which has various forms as well as easy, fast and wide spread, then the effect is felt increasingly significant. So it can be said that digital humor can be used as a coping strategy for stress in the work environment.

However, not all research presents positive results, the use of humor that is not appropriate, such as making yourself or others as a joke can make others have negative views and reduce social support, so that it can impact on higher stress levels According to Markman (2017) . In addition, several studies have shown that the use of humor as a coping strategy is an ineffective way to deal with stress, so that its use when compared with other coping strategies including low frequency (Gutierrez, et al, 2007; Washburn-Ormachea, et al, 2004; Wu & Chan, 2013). The use of humor as a stress management tool is not universal, but depends on the culture that is adopted, so individuals must continue to maintain behavior in accordance with polite and polite ethics (Yue, 2010).

The existence of this negative research result does not cover the fact that digital humor can be used as a stress coping strategy with its certainly taking into account the norms and ethics that apply in a community or entity. Being wise in using digital humor is very important, because if not, with its distribution that is easy, fast and broad it will present new problems which will lead to rising stress levels. Further research that examines the effect of digital humor on work stress even further on performance is still very much needed because it has not been found in previous studies.

**References**
Ahmadi, D. 2005. Interaksi Simbolik: Suatu Pengantar. Mediator, 9 (2): 301-314.
Ardianto, E., Komala, L., & Karlinah, S. 2007. Komunikasi Massa Suatu Pengantar, Revisi. Bandung: Simbiosa Rekatama Media.
Bondaroux, T, H R., & Ruel, B, V D. 2009. E-HRM Effectiveness in a Public Sector Organization: A Multi-Stakeholder Perspective. The International Journal of Human Resources Management. 20 (3). 578-590

Budiawati, H.. 2016. Identifikasi Sumber Stres Tenaga Pengajar dan Pengaruhnya terhadap Kinerja Dosen di STIE Widya Gama Lumajang. Jurnal Penelitian Ilmu Ekonomi WIGA, 6 (1): 27-35.

Bungin, B. 2004. Metode Penelitian Kualitatif. Jakarta: PT Rajagrafindo Persada

Catanescu, C. & Tom, G. 2001. Types of Humor in Television and Magazine Advertising. Review of Business. Summer 2001.

Cohen, S., & Williamson, G. 1988. Perceived Stress in a Probability Sample of the United States. Spacapan, S. and Oskamp, S. (Eds.) The Social Psychology of Health. Newbury Park, CA: Sage.

Colom, G.G., Alcover, C.T., Curto, C.S., & Osuna, J.Z. 2011. Study of The Effect Of Positive Humour As A Variable That Reduces Stress. Relationship of Humour With Personality And Performance Variables. Psychology in Spain, 15 (1): 9-21.

Cooper, D. R. and Schindler, P. S. 2014. Business Research Methods. Boston: McGraw-Hill International Edition.

Doosje, S., De Goede, M., Van Doornen, L., & Goldstein, J. 2010. Measurement of occupational humorous coping. Humor, 2 (3), 275–305.

Ghozali, Imam. 2002. Aplikasi Analisis Multivariate dengan program SPSS, Semarang: Badan Penerbit Universitas Diponegoro

Gibson, I., & Donnelly. 1993. Organisasi dan Manajemen: Perilaku, Struktur, Proses. Jakarta: Erlangga.

Gutiérrez, F., Peri, J. M., Torres, X., Caseras, X., & Valdés, M. 2007. Three dimensions of coping and a look at their evolutionary origin. Journal of Research in Personality, 41(5), 1032-1053.

Hamidi. 2004. Metode Penelitian Kualitatif. Malang: UMM Press.

Hariyono, W., Suryani, D., & Wulandari, Y. 2004. Hubungan antara Beban Kerja, Stres Kerja, Dan Tingkat Konflik Dengan Kelelahan Kerja Perawat Di Rumah Sakit Islam Yogyakarta Pdhi Kota Yogyakarta. Jurnal Kesmas, 3(3): 162-232.

Luthans, F. 2011. Organizational behavior. New York: McGraw Hill.

Mangkunegara, A.P. 2008. Perilaku dan Manajemen Organisasi. Jakarta: PT. Erlangga.

Markman, A. 2017. Humor Sometimes Makes Stressful Situations Better. Psychology Today. Retrieved from http://www.psychologytoday.com

Martin, R. A. 2007. The Psychology of Humor: An Integrative Approach. London: Elsevier

Martin, R.A., Doris, P., Larsen, G.G., & Weir, K. 2003. Individual differences in uses of humor and their relation to psychological well-being: Development of the Humor Styles Questionnaire. Journal of Research in Personality, 37, 48–75.

Rahayu, E., & Handriani, E. 2015. Stres dan Sense of Humor pada Guru SLB C. Psikodimensia, 14 (2): 41-54.

Rahmansari, R. 2017. Penggunaan aplikasi whatsapp dalam komunikasi organisasi pegawai dinas lingkungan hidup dan kebersihan Sidoarjo. Jurnal ilmiah manajemen publik dan kebijakan sosial, 1(2): 77-90.

Ramya, U., & Poornachandran, B. 2017. A Study On Impact Of Stress On Quality Of Work Life Among Women Employees (With Special Reference To Chennai City). International Journal of Latest Engineering and Management Research, 2(11): 45-49.

Raskin, V. 1985. Semantic Mechanisms of Humor. Reidel Publishing. Dordrecht, the Netherlands.

Reyes, A., Rosso, P. & Buscaldi, D. 2012. "From humor recognition to irony detection: The figurative language of social media". Data & Knowledge Engineering. 74. pp. 1-12.

Robbins, S. 2012. Perilaku Organisasi. Jakarta: Penerbit Salemba Empat.

Robbins, S.P, & Judge T.A. 2015. Perilaku Organisasi. Jakarta: Salemba Empat.

Shifman, L. 2007. Humor in the age of digital reproduction: continuity and change in internet-based comic texts. International Journal of Communication, 1: 187-209.

Silalahi, U. 2009. Metode Penelitian Sosial. Bandung: PT. Refika Aditama.

Siswanto. 2007. Kesehatan Mental: Konsep, Cakupan dan Perkembangannya. Yogyakarta: Andi Offset.

Sobur, A. 2004. Semiotika Komunikasi. Bandung: Rosda Karya.

Sugiyono. 2012. Metode Penelitian Kuantitatif Kualitatif dan R&D. Bandung: Alfabeta

Techarungroj, V., & Mueang jamnong, P. 2014. The effect f humour on virality: the study of internet memes on social media. Paper presented on Forum of Public Relations.

Torretta, A. 2014. A Funny Thing Happened on the Way to the Fair: Using Humor to Decrease Stress and Increasing Productivity. Journal of Extension, 52 (3). 1-17

Tümkaya, S. 2007. Burnout and humor relationship among university lecturers. Humor, 20 (1), 73-92.

Ulber, Silalahi. 2009. Metode Penelitian Sosial. Bandung: PT. Refika Aditama

Washburn-Ormachea, J. M., Hillman, S. B., & Sawilowsky, S. S. (2004). Gender and gender-role orientation differences on adolescents' coping with peer stressors. Journal of Youth and Adolescence, 33 (1), 31-40.

Widyowati, A., & Priambodo, E.P. 2016. Hubungan antara Kepekaan Humor Dengan Stres Kerja Pada Pegawai Balai Karantina Pertanian Kelas 1 Semarang Jawa Tengah. Psikologika, 21(1): 47-56.

Wijaya, E. 2017. Peranan Humor Terhadap Stres Dengan Subjective Well Being (SWB) Sebagai Mediator Pada Dewasa Awal. Jurnal Muara Ilmu Sosial, Humaniora, dan Seni, 1 (1): 353-360.

Wirawan. 2015. Manajemen Sumber Daya Manusia Indonesia. Jakarta: Rajawali Press.

Wu, J., & Chan, R. 2013. Chinese teachers' use of humour in coping with stress. International Journal of Psychology, 48 (6), 1050-1056.

Yin, R, K. 2014. Studi Kasus Desain & Metode. Jakarta: Rajawali Pers.